\documentstyle[12pt]{article}
\normalbaselines
\begin{document}
\begin{center}
\vspace*{1.0cm}
{\LARGE \bf \baselineskip2.8ex
 Problems about Causality in Fermi's  Two-Atom Model
 and Possible Resolutions\footnote
 {This article has appeared in:
  NONLINEAR, DEFORMED AND IRREVERSIBLE QUANTUM SYSTEMS,
  Proceedings of an International Symposium on Mathematical Physics at the
  Arnold Sommerfeld Institute 15--19 August 1994, Clausthal, Germany.
  Editors: H.-D. Doebner, V.K. Dobrev, P. Nattermann.
  WORLD SCIENTIFIC, Singapore (1995), p. 253 -- 264}\par }
\vskip 1.5cm
{\large \bf  Gerhard C. Hegerfeldt  }\vskip 0.5 cm
Institut f\"ur Theoretische Physik\\
Universit\"at G\"ottingen \\
Bunsenstr. 9\\
D-37073 G\"ottingen
\end{center}
\vspace{1 cm}
\begin{abstract}
In order to check finite propagation speed
Fermi, in 1932, had considered two atoms $A$ and $B$ separated by some
distance $R$. At time $t = 0$, $A$ is in an excited state, $B$ in its
ground state, and no photons are present. Fermi's idea was to
calculate the excitation probability of $B$. In a model-independent
way and with minimal assumptions -- Hilbert space and positive energy
only -- it is proved, not just for atoms but for any systems $A$ and
$B$, that the excitation probability of $B$ is nonzero immediately
after $t = 0$. Possible ways out to avoid a contradiction to finite
propagation speed are discussed. The notions of strong and weak
Einstein causality are introduced.
\end{abstract}
\vspace{1 cm}
\section{Introduction}

One of the pillars of special as well as general
relativity is the assumption that no signals can be transmitted
faster than the speed of light. If there were arbitrarily high signal
velocities in nature then either
\begin{itemize}
\item those superfast signals could be used to synchronized clocks to
yield absolute simultaneity and thus a breakdown of relativity
theory, or
   \item there would exist ``tachyons", and the sequence of cause and
effect could be reversed.
\end{itemize}

The second alternative has captured imaginative minds and prompted 
them to create science-fiction like scenarios. The concept of finite
signal velocity or, more precisely, the speed of light as highest
signal velocity, is therefore often called ``Einstein causality". 
In my opinion, though, if Einstein causality were to fail most
physicist would adopt the first alternative and reformulate or
abandon relativity theory. 

For this reason the question of finite signal velocity in quantum
theory attracted the interest of Heisenberg and Fermi in the early
thirties, in particular whether photons traveled with the speed of
light.

In 1932, Fermi \cite{1} consider for this purpose a simple model.
Two atoms, $A$ and $B$, are separated by a distance $R$. At time $t =
0$, $A$ is assumed to be in excited state and $B$ in its ground
state, with no photons present. Atom $A$ will  
decay into its ground state under the emission of a photon.
This photon can then, with a small probability, be absorbed by atom
$B$. Fermi asked the question at what earliest time atom $B$ will
``notice" the decay of atom $A$ with its accompanying photon. 
He expected that $B$ moves out of its ground state only after a time $t =
R/c$, in accordance with the speed of light. And indeed, this was
what he found by his calculations.

Fermi's calculations were based on second-order perturbation theory
-- a technique still quite common today -- and on the approximation
of an integral over positive frequencies by an integral over positive
and negative frequencies ranging from $- \infty$ to $\infty$ instead
of $0$ to $\infty$. More than thirty years later Shirokov \cite{2}
pointed out that without this replacement the calculations would not
yield the desired result \cite{Nature}. It remained unclear, however,
what would happen if one went to higher orders in perturbation
theory.

The setup of Fermi's model will be further discussed in the next
section. Fermi had calculated the probability for the following
transition: $A$ nonexcited, $B$ excited and no photons. As will be
discussed this is an exchange probability \cite{2} and it does not
directly, without further assumptions, refer to Einstein causality
but to what one nowadays calls local and nonlocal correlations.

Fermi's problem was investigated by many authors in this or in a
related form, e.g.  by Heitler and Ma \cite{HeiMa}, Hamilton
\cite{Ham}, Fierz \cite{Fierz}, Ferretti \cite{Fer}, Milonni and
Knight \cite{MiKni}, Shirokov \cite{2} and his review \cite{Shir2},
Rubin \cite{Rub}, Biswas et al. \cite{Bis}, and Valentini \cite{Val}.
The older papers confirmed Fermi's conclusion, while the results of
the later papers depend on the model and the approximations used. At
present there seems to be agreement that Fermi's `local' result is not correct,
but that this nonlocality cannot be used for superluminal signal
transmission since measurements on $A$ and $B$ as well as on photons are
involved. 

The present contribution, which is partially based on Ref.
\cite{self}, 
is mathematically very simple. It analyzes Fermi's model under quite
general and simple assumptions -- essentially just positivity of the
energy. No perturbation theory, no specific form of the Hamiltonian,
nor further assumptions of quantum field theories like the locality
postulate, are used, and the conclusions hold for relativistic and
nonrelativistic theories. Moreover, the atoms can be replaced by more
general ``sources" and ``detectors".

Specifically, it is shown for the model  considered by Fermi that the
excitation probability of atom $B$ would be immediately nonzero if
the experiment could really be performed. At first sight this result
might seem to indicate serious difficulties with causality for
Fermi's two-atom model. However, already in Ref. \cite{self} I
pointed out several ways to avoid this disastrous consequence, and in
the last section I will discuss additional ones. The message is that
finite signal velocity is a delicate question.

Somewhat surprisingly, the results of my paper \cite{self} or received
great publicity and  were 
discussed not only in science journals like Nature \cite{Nature} or
New Scientist \cite{NS}, but also made it to the daily press and
weekly magazines \cite{press}. While some discussions were reasonably
serious, smaller tabloids tended to sensationalize by picking on
acausality and omitting the ways out \cite{Go}. The moral to draw from
this is that one should not rely on second or third hand accounts,
in particular not on sensational ones.

\section {Fermi's model: Correlations, Excitation
Probabilities, and Bare States}

\noindent Fermi supposed in his model that by some means one had prepared, at time
$t = 0$, atom $A$ in an excited state, $| e_A \rangle$, and atom $B$
in its ground state, $|g_B \rangle$, with no photons present. The
state of the complete system then developed in time and Fermi
calculated the ``exchange" probability to find the state $|g_A
\rangle\!|e_B \rangle\!|0_{ph} \rangle$ at time $t$. He
probably had in mind that this could occur only by deexcitation of
$A$, emission of a photon by $A$, absorption of it by $B$ and
excitation of $B$. However, actually to check that there are no
photons requires, at least in principle, photon measurement over all
space, not only measurements of the states of $A$ and $B$. Hence such
an exchange probability cannot be used for signals, it just refers to
statistical correlations. Really needed in this model approach to
finite signal velocity is the probability of finding $B$ excited,
irrespective of the state of $A$ and possible photons; if there turn
out to be no photons, all the better. This excitation probability
could then, in Fermi's approach, be determined by a measurement on $B$
alone. 

Using ``bare" states, as Fermi did, the Hilbert space is simply the
tensor product
\begin{equation}\label{tensor}
{\cal H}_{\rm bare}~=~{\cal H}_A\times
{\cal H}_B\times{\cal H}_F
\end{equation}

\vspace*{0.5cm}

\noindent and the Hamiltonian is of the form

\vspace*{0.5cm}

\begin{equation}\label{Hbare}
H_{\rm bare}~=~H_A~+~H_B~+~H_F~+~H_{A F}~+~H_{B F}~.
\end{equation}
At time $t = 0$, the initial state is
\begin{equation}\label{bstate}
| \psi_0{}^{\rm bare} \rangle~=~ |e_A \rangle |g_B \rangle |0_{ph} \rangle~.
\end{equation}
At time $t$, the probability of finding $B$ in some excited state is
then a sum overall excited states $| e_B \rangle$, over all states
$| i_A \rangle$ of $A$ and over all photon states $| \{ {\bf n}\}
\rangle$, i.e.
\begin{eqnarray}
\lefteqn{\sum_{e_B} \sum_{i_A}  \sum_{\{{\bf n}\}}| \langle \{ {\bf n} \}|
\langle e_B | \langle i_A| \psi_t^{\rm bare} \rangle |^2} \nonumber
\\
& = & \langle \psi_t^{{\rm bare}}|~\{\sum_{i_A, e_B, \{{\bf n}\}}|i_A
\rangle | e_B\rangle
|\{ {\bf n}\} \rangle \langle \{
{\bf n}\} | \langle e_B |\langle i_A |\}~|\psi_t^{{\rm bare}} 
\rangle \nonumber\\
& = &   \langle \psi_t^{{\rm bare}}|~ {\bf 1}_A\times
\sum_{e_B}|e_B \rangle \langle e_B  |\times {\bf 1}_F~
| \psi_t^{{\rm bare}} \rangle \label{exc}~.
\end{eqnarray}
The r.h.s. is the expectation of the operator
\begin{equation}\label{bobs}
{\cal O}^{{\rm bare}}_{e_B}~ \equiv~ {\bf 1}_A \times 
\sum_{e_B}|e_B \rangle \langle e_B |
\times {\bf 1}_F~.
\end{equation}
This operator represents the observable ``$B$ is in a bare excited
state", and here it is a projection operator.

\section{ Renormalized States}

Bare states are widely used in quantum optics and are
usually quite adequate. However, for subtle questions of principle of
a physical theory great care is needed. Approximations and
perturbation theory may give misleading results. In one order an
effect might show up, but not in the next order, and so on.
Unrenormalized bare theories are, without cut-off, mathematically not
well-defined. They are plagued by infinities whose cancellation has
not been investigated for signal velocities.

For the present purpose it suffices to use only rudiments of a
renormalized theory. We just need the following two simple
properties,
\begin{itemize}
   \item[(i)] existence, with a Hilbert space ${\cal H}_{\rm ren}$, 
   \item[(ii)] a Hamiltonian, $H_{{\rm ren}}$, which is bounded from
below and self-adjoint (``positive energy").
\end{itemize}
Then, in general, ${\cal H}_{\rm ren}$ is no longer the tensor
product in Eq. (\ref{tensor}), and the initial state, denoted by $|
\psi_0 \rangle$, will not be a simple product state,
\[
|\psi_0 \rangle~\not=~|e_A \rangle | g_B \rangle |
0_{\rm ph} \rangle~.
\]
Similarly, if the observable
\begin{equation}\label{6a}
\mbox{``B is in an excited state"}
\end{equation}
makes sense and is represented by an operator $O_{e_B}$ then in
general
\[
{\cal O}_{e_B}~\not=~{\cal O}^{\rm bare}_{e_B}~.
\]
However, its expectation values must lie between $0$ and $1$ to
represent a probability, i.e., 
\begin{equation}\label{7a}
0 \le {\cal O}_{e_B} \le 1~.
\end{equation}
For example, ${\cal O}_{e_B}$ might be a projector, as for bare states, but
this would not be the most general case. Thus 
\[
P^e_B (t) \equiv \langle \psi_t |{\cal O}_{e_B}\!|\psi_t \rangle
\]
would be the excitation probability of $B$ at time $t$, and it would
involve a measurement on $B$ only. It should be noted that the
explicit form of the operator ${\cal O}_{e_B}$ is not required in the
following, only Eq. (\ref{7a}) will be used.

Also no point-like localization of $A$ and $B$ are required.
Generalizing Fermi's model, $A$ and $B$ may be systems initially
localized in two regions separated by a distance $R$, 
with no photons present. The ground state of $B$ may be degenerate.
Again, with Fermi, one would suppose that one had somehow managed to
prepare this initial state at $t = 0$. The analog of Fermi's original
result would then be that 
the excitation probability $P^e_B (t)$ of $B$ would vanish for $t \le
R/c$.

In the next section I will prove a simple mathematical theorem which
applies to this situation and which yields that either
\begin{itemize}
   \item[(i)] $~~~~P^e_B (t) \not= 0~~~~~$ for almost all $t$,
   \end{itemize}
or
\begin{itemize}
   \item[(ii)] $~~~~P^e_B (t) \equiv 0~~~~~$  for all $t$~.
\end{itemize}
This result does not agree with Fermi's original expectation . 
In the last section I will show how one can by-pass potential
difficulties for finite signal velocity 
by modifying and clarifying the {\it physical} assumptions employed.
There is also a mathematical loophole which could be used, although
the theorem, of course, remains true.

\section{ The Theorem}

To more clearly separate what is Physics and what
Mathematics, I will phrase the theorem in purely mathematical terms
although its main and possibly sole interest lies in its
applications to the physical situation described above. So what was
previously $H_{\rm ren}$ now becomes any self-adjoint operator $H$
bounded from below, and the initial state $|\psi_0 \rangle$ can
now be any state, while in the application it represents a physical
situation in which $A$ is supposed to be in an excited state, $B$ in
a ground state and with no photons. 

\vspace*{0.5cm}

\noindent {\bf Theorem.} Let $H$ be self-adjoint and bounded from
below and let ${\cal O}$ be any operator satisfying
\begin{equation}\label{8a}
0 \le {\cal O} \le 1~.
\end{equation}
Let $\psi_0$ be any vector and define
\[
\psi_t \equiv e^{-i Ht/\hbar} \psi_0~.
\]
Then one of the following two alternatives hold.
\begin{itemize}
   \item[(i)] $~~~~\langle \psi_t , {\cal O}~ \psi_t \rangle \not=
0~~~~~~$ for almost
all $t$, and the set of such $t$'s is dense and open.
\item[(ii)] $~~~~\langle \psi_t,{\cal O}~\psi_t \rangle \equiv 0~~~~~~$
for all $t$.
\end{itemize}

\vspace*{0.5cm}

\noindent {\bf Proof.} Let us define 
\begin{equation}\label{9a}
P (t) = \langle \psi_t,{\cal O}~\psi_t \rangle~.
\end{equation}
Since $\psi_t$ is continuous in $t$, so is $P(t)$. From
this it follows immediately that the set ${\cal N}_0 := \{t ; P(t) = 0\}$ is
closed and its complement ${\cal N}^c_0$ is open. Since ${\cal O}$ is a
positive operator, its positive square-root ${\cal O}^{1/2}$ exists, and one
has 
\[
\langle \psi_t,{\cal O}~\psi_t \rangle = \langle {\cal O}^{1/2} \psi_t, 
{\cal O}^{1/2}
\psi_t \rangle~.
\]
For $t \in  {\cal N}_0$ this vanishes, and thus 
\begin{equation}\label{12}
{\cal O}^{1/2} \psi_t = 0~~~~~~ \mbox{for}~ t \in {\cal N}_0~.
\end{equation}
Now let $\phi$ be any fixed vector and define the auxiliary function
$F_\phi (t)$ by 
\begin{equation}\label{13}
F_\phi (t) = \langle \phi,{\cal O}e^{- i H t/\hbar} \psi_0 \rangle.
\end{equation}
Hence, by Eq. (\ref{12}), 
\begin{equation}\label{13a} 
F_\phi (t) = 0 ~~~~~~\mbox{for}~t \in {\cal N}_0~.
\end{equation}
Since $H \ge - ~\mbox{const}$, one has that the operator
\[
e^{- iH (t + i y)/\hbar}
\]
is well-defined for $y \le 0$. Putting
\[
z = t + i y
\]
one sees that $F_\phi (z)$ can be defined as a continuous function
for Im $z \le 0$, and, moreover, $F_\phi (z)$ is analytic for Im $z <
0$. 

Let us now assume that (i) does not hold, i.e. that either ${\cal
N}_0$ is not a null set or that its complement ${\cal N}_0^c$ is not 
dense. It would suffice to consider the former, but the
latter can be treated in an almost elementary way, so I consider it
first. If ${\cal N}_0^c$ is not dense, ${\cal N}_0$ contains some
nontrivial interval, $I$ say. Hence $F_\phi (z)$ vanishes on $I$, by
Eq. (\ref{13a}), and one can directly use the Schwarz reflexion
principle \cite{22} or proceed as follows. One defines an extension
of $F_\phi$ to the upper half plane by putting
\begin{equation}\label{14}
F_\phi (z) = F_\phi (z^*)^*~~~\mbox{for}~ \mbox{Im}~z > 0~.
\end{equation}
Since $F_\phi (t)$ is real for $t \in I$ it follows that the extension is
continuous on $I$, and from this one can show that it is analytic for
$z \not\in I\!\!R \backslash I$. Hence $I$ is contained in the
analyticity domain. Since $F_\phi (z) = 0$ for $z \in I$, it
therefore vanishes
identically in its domain of analyticity. However, since it is
continuous when approaching the real axis, it follows that 
$F_\phi (t) = 0$ for all
$t$. Since $\phi$ was arbitrary this implies 
\[
{\cal O}\psi_t = 0~~~\mbox{for all}~ t,
\]
and this gives case (ii).

This proves the interesting part of the theorem, namely that $P (t)$
is either nonzero on a dense open set or that it vanishes
identically. Since a dense open set need not have full Lebesgue
measure this does not yet prove the full theorem. However, as a
boundary value of a bounded analytic function, $F_\phi (t)$ satisfies
the inequality \cite{23}
\begin{equation}\label{15}
\int^\infty_{- \infty} dt \ln | F_\phi (t) | / (1 + t^2) > -
\infty
\end{equation}
unless it vanishes identically. If ${\cal N}_0$ had positive measure
the integral would be $- \infty$, and thus $F_\phi (t)$ would vanish
identically, for each $\phi$. This would again imply case (ii).
Incidentally, this last argument also covers the previous case since,
if ${\cal N}_0$ is a null set, its complement is dense. Since the
argument based on the Schwarz reflexion principle is very transparent
it has been included. This completes the proof of the theorem.

There is a similarity of this result with the Reeh-Schlieder theorem
\cite{RS} which also exploits analyticity but uses stronger
assumptions of field theory, in particular locality. It is therefore
not directly applicable to the general situation considered here.

Taking for ${\cal O}$ in the theorem the previously considered observable
${\cal O}_{e_B}$ and for $\psi_0$ the state $|\psi_0 \rangle$
representing the initial state with $A$ excited, $B$ in its ground
state and no photons -- provided they exist -- one obtains from the
theorem that the excitation probability of $B$ is immediately nonzero
after $t = 0$ -- unless it vanishes for all times, a case one might
exclude on physical grounds.

Another application can be made to the correlations mentioned in the
Introduction. Let $|\psi_{\rm ex} \rangle$ denote the state
representing $A$ in a ground state, $B$ in an excited state, and no
photons, either in ${\cal H}_{\rm ren}$ or ${\cal H}_{\rm bare}$,
provided again the notion makes sense in the case of ${\cal H}_{\rm
ren}$. In the bare case one just has
\[
|\psi_{\rm ex}\!\rangle_{\rm bare} = |g_A \rangle |
e_B \rangle|0_{\rm ph} \rangle~.
\]
We define
\[
{\cal O}_{\rm ex} \equiv~ |\psi_{\rm ex} \rangle \langle \psi_{\rm ex}
|~.
\]
The expectation value of ${\cal O}_{\rm ex}$, 
\[
\langle \psi_t |{\cal O}_{\rm ex}|\psi_t \rangle = |\langle
\psi_{\rm ex}|\psi_t \rangle |^2
\]
is just the transition probability to $|\psi_{\rm ex} \rangle$.
Since ${\cal O}_{\rm ex}$ is a projector the theorem yields that the
transition probability is immediately nonzero, unless it vanishes
identically. But since this is just a correlation function of
measurements at different positions this result has no bearing on
signal velocities.

Further, more general, applications are possible. $A$ and $B$ can be
any quantum systems, e.g. a ``source" and detector; they may be
moving. One may also envisage other particles and interactions. One
may apply it also to a problem of Heisenberg who had suggested  to
consider an excited atom $A$ with no photons and to calculate the
probability to find a photon at time $t$ in a region a distance $R$
away. At that time this probability was found to vanish for 
$t < R /c$ \cite{26}. 

\vspace*{1cm}

\section{ Discussion and Ways out}

\vspace*{0.5cm}

\noindent As already stressed in Ref. \cite{self}, the theorem is
a mathematically rigorous result, and its applications in Physics
depend on the physical assumptions, leading to statements of the
form, ``If ..., then ...". For example, I had been careful, when
introducing the excitation observable ${\cal O}_{e_B}$, to say, ``if this
makes sense". If it does exist, then indeed the excitation
probability of B is immediately nonzero, or identically zero which
one would exclude. Does this mean that atom $B$ has been
excited by a superluminal photon emitted by $A$? Not necessarily.
Before discussing this we discuss another, more mathematical,
way out.

\vspace*{0.6cm}

\noindent {\it Possible mathematical resolution}

\vspace*{0.4cm}

An {\it explicit} calculation of transition probabilities or other
quantities in quantum optics, or quantum electrodynamics or theories
involving fields, will in general not start from a renormalized
theory with Hilbert space ${\cal H}_{\rm ren}$ and Hamiltonian 
$H_{\rm ren}$ -- the
form is not known, not even the existence. Instead one will introduce
cutoffs in the bare theory to make it well-defined, then calculate
transition probabilities, and finally one will remove the cutoffs,
taking care of divergent expressions by renormalization. For
each cutoff the theorem may be applicable and may yield a nonzero
probability for almost all $t$ in $t < R/c$. However, as the cutoffs
are tending to infinity, the nonzero probability in this time
interval may in principle become smaller and smaller, and in the limit
one might conceivably have $0$ for $t < R/c$. If this were so, then
the mathematical assumptions -- existence of ${\cal H}_{\rm ren}$,
$ H_{\rm ren}$, and ${\cal O}_{e_B}$ -- could not be fulfilled. Such a
possibility, nonexistence of a Hilbert space after renormalization,
has indeed been discussed in the literature \cite{Fr}. 

\vspace*{0.6cm}

\noindent {\it  Physical ways out} 
\vspace*{0.4cm}

We now discuss the more or less implicit
physical assumptions that have been made,  and how to by-pass them.

\vspace*{0.4cm}

\noindent {\it (a)} Systems localized in disjoint regions might not
exist as a matter of principle, i.e. systems might always ``overlap".

In ordinary quantum mechanics the wave-function associated with an
energy level of a hydrogen atom extends to infinity, and this has
been proposed before as a reason for overlapping \cite{Mad}. But
since this happens in a nonrelativistic theory, this particular
argument probably does not go to the heart of the matter. Moreover,
in nonrelativistic quantum mechanics, there do exist wave-functions
of the hydrogen atom which vanish outside some finite volume at a
given fixed time. By completeness these wave-function can be obtained
by suitable superpositions, but they spread out to infinitely
instantaneously \cite{HeRu}.

A better argument for overlapping may seem that one might conceivably
create particle-antiparticle pairs or other particles whenever one
tries to localize a system too well. This has been advanced as an
explanation for the difficulties one has in obtaining good
localization or position operators \cite{13,14,15,17,18}. This is
connected to the next point.

\vspace*{0.5cm}

\noindent {\it (b)} Renormalization may introduce a sort of photon
cloud around each system, e.g. due to "vacuum fluctuations".
This essentially means an overlapping of
the systems with their clouds and leads back to (a). More
specifically, photons of one cloud may excite one or the other
system, and this might even happen with only one system present.

\vspace*{0.5cm}

\noindent {\it (c)} The notion of a ground state of $B$, either with
or without $A$ present, might not make sense, due to renormalization.  

\vspace*{0.6cm}

\noindent {\it Strong and weak Einstein causality}

\vspace*{0.4cm}

\noindent How then to check finite propagation speed? To clarify matters it is
useful to differentiate between two notions of Einstein causality.

\vspace*{0.4cm}

\noindent {\it (i) Strong Causality}: For {\em each} individual
process or experiment  there is no excitation or disturbance of the
second system for $t < R/c$.

This notion is similar to energy -- momentum and Baryon conservation
in each individual scattering process in particle physics. Strong
causality would hold if the transition probabilities considered above
were strictly zero for $t < R/c$. It seems to me that both Fermi
\cite{1} and Heisenberg - Kikuchi \cite{26} had this in mind when
they set out to prove that certain probabilities vanished for $t <
R/c$. The above theorem shows that strong causality cannot be {\em
checked}, unless the way out via cut-off theories holds, or it may
fail, a possibility I do not advocate.

\vspace*{0.4cm}

\noindent {\it (ii) Weak Causality}: This notion was introduced in
Ref. \cite{Schl}, and loosely speaking it means that Einstein
causality holds for expectation values only, i.e. for {\em
ensembles}, not for individual processes. For weak causality to hold,
expectation values, i.e. ensemble averages, need not vanish for $t <
R/c$, but it takes a time at least $t = R/c$ to produce an effect on
them. To exhibit this effect one may suitably subtract possible
fluctuations of system $B$ alone, e.g. vacuum fluctuations.

Without additional assumptions to those of the theorem (Hilbert
space, positive energy) probably nothing can be said about weak
causality in Fermi's setup. Model calculations \cite{Bis,Val} point in
the right direction, although the renormalization problem remains
unsolved. ``Bare" theories can sometimes give useful  indications on
the problem of weak causality, but cannot provide definitive answers.
This is all the more true for bare theories with nonrelativistic
atoms.

Buchholz and Yngvason \cite{R6} use the assumptions of the
theory of local observables (``algebraic quantum field theory"),
which are stronger than those employed for the above theorem. They
point out that in this theory transition probabilities and the above
observable ${\cal O}_{e_B}$ are no legitimate quantities and thus not
allowed. The idea is therefore to consider only
observables which are allowed in the theory of local observables.
These are observables associated with bounded space-time regions; in
particular, sharp-time observables are excluded. To give the essential
ideas, let $| \psi_{A B} (t) \rangle$ denote the state with
systems $A$ and $B$ present, $| \psi_B (t) \rangle$ the state with
only $B$ present, and let ${\cal B}$ be an observable associated with a
space-time region of system $B$. Then weak causality requires 
\[
\langle \psi_{A B}(t) | {\cal B} | \psi_{A B}(t) \rangle =
\langle \psi_{B}(t) | {\cal B} | \psi_{B}(t) \rangle
~~~~~~~\mbox{for}~~ t < R/c~;
\]
i.e. only the difference of both sides is zero for $t < R/c$. With
the assumptions inherent in the theory of local observables this does
hold \cite{R6}. 

For simplicity let us assume that this could be applied to Fermi's
question about the excitation probability of $B$ (as remarked before,
this is not a good observable in algebraic quantum field theory).
This would then mean that the excitation probability of $B$ could
indeed be nonzero, but until $t = R/c$ it would not depend on the
presence of $A$. 

What does this mean physically? Expectation values
are ensemble averages, and to check this nondependence on $A$
experimentally one could not use a {\em single} pair of systems $A$
and $B$. Instead one would need an ensemble of such pairs -- either
by repetition of the experiment or by simultaneous realization of
many, $N$ say, of such pairs at $t = 0$. At time $t$ one would then
measure how many of the $B$ systems are excited. Their fraction is
\[
 P^e_{B {\rm (with~ A)}}(t) = P^e_B (t)
\]
if $N \rightarrow \infty$; for finite $N$ this holds only
approximately, due to statistical fluctuations. Now one would
calculate the excitation probability of $B$ without system $A$
present, which is denoted by $P^e_{B {\rm(w/o~A)}}(t)$, and subtract
it. Weak causality would then assert
\begin{equation}\label{w}
P^e_{B{\rm (with~ A)}}(t) - P^e_{B {\rm (w/o~ A)}}(t) = 0
~~~~\mbox{for}~ t < R/c~ .
\end{equation}
Only for $N \rightarrow \infty$ would this be strictly true experimentally
since for finite $N$ there would always be statistical fluctuations.

Hence finite propagation velocity or speed of light in the sense of
weak causality cannot be checked experimentally in a  strict
sense for {\em finite} ensembles since in this case there are always
deviations from the exact zero in Eq. (\ref{w}). What is needed are
rigorous bounds on the $N$ dependence of the statistical fluctuations.
The theory of local observables does not provide these, and rigorous
model studies may be more promising for the question of bounds.

In a strict sense, finite propagation speed as expressed by weak
Einstein causality can only be checked experimentally for {\em
infinite} ensembles, and this may suggest that this notion 
somehow belongs to a macroscopic context.


\begin{thebibliography}{99}
\bibitem{Go} At least one local paper even had a front page headline:
``There are things faster than light". One can imagine the weird 
calls I got.
\end{thebibliography}
\end{document}